\documentstyle[epsf,aps]{revtex}
\tighten
\begin{document}
\draft
\title{Non-equilibrium phase transition in a two-temperature lattice gas}
\author{Attila Szolnoki}
\address
{Research Institute for Materials Science, H-1525 Budapest, POB 49, Hungary}
\maketitle                                                           
\begin{abstract}

A two-temperature lattice gas 
model with repulsive nearest-neighbour interactions is studied 
using Monte Carlo simulations and dynamical mean-field approximation. 
The evolution of the two-dimensional, half-filled system is 
described by an anisotropic Kawasaki dynamics assuming that the
hopping of particles along the principal directions is
governed by two heat baths at different temperatures
$T_x$ and $T_y$.
The system undergoes an order-disorder phase transition as $T_x$ 
($T_y$) is varied for sufficiently low fixed $T_y$ ($T_x$).
The non-equilibrium phase transition remains continuous and 
the critical behaviour belongs to the Ising universality
class. 
The measure of violation of the fluctuation-dissipation 
theorem can be
controlled by the value of the fixed temperature. 
We have found an exponential decay of spatial correlations
above the critical region in contrast to the two-temperature 
model with attractive interactions.   
\end{abstract}
\pacs{05.50.+q, 05.70.Ln, 64.60.Cn}

\section{Introduction}
The study of driven diffusive lattice gas model, which served 
a convenient tool to produce 
stationary non-equilibrium state, revealed a lot of new phenomena
compared to the 
corresponding equilibrium properties \cite{zia}. Such a phenomenon 
is the emergence of a novel 
non-equilibrium fixed point of critical behaviour in the case of 
attractive interactions or 
the surprising feature of the fluctuations of internal energy \cite{jans,vm}. 
In this model an external uniform electric field is applied to induce a
preferential hopping in one direction leading to a permanent particle 
transport in the stationary state \cite{kls}. 
Unfortunately, it is 
rather difficult to realise the model in laboratory because the driving field 
cannot be derived as the gradient of an electrostatic potential. 

This problem can be avoided if we use an alternating field instead of the
uniform one, which 
also enlarges the probability of jumps along specified lattice directions 
without resulting in a non-zero average particle current. 
Similar effect can be obtained in a two-temperature model where 
the hopping of particles along the principal directions is governed by two 
heat baths at different temperatures \cite{garr,cheng}. Now the flow
of energy between the heat baths drives the system into a non-equilibrium 
stationary state. The non-equilibrium phase diagrams have already been 
studied in two-dimensions in the case of attractive interactions 
\cite{racz} and for nearest-, and next-nearest neighbour repulsive 
interactions \cite{sso}. Common feature of the two diagrams is the 
existence of an ordered phase even at $T_x = \infty$, if $T_y$ is low enough. 
Similar behaviour is observed when the uniform electric 
field drives the systems into non-equilibrium states. Here the ordered 
phases also exist at infinite E if the temperature is low enough 
\cite{kls,ss}. Completely different behaviour is found for 
nearest-neighbour repulsive interactions
where the field induced interfacial transport prevents the long-range 
ordering \cite{ssa}. Although similar effect cannot be expected for the 
corresponding two-temperature model because of the zero 
macroscopic current, still the stability of the ordered phase is doubtful 
\cite{zia}.

In this paper we determine the phase diagram of a two dimensional 
two-temperature model with nearest-neighbour repulsive interactions. 
For this purpose we perform dynamical mean-field 
analysis to calculate the phase transition points. These results are 
supported by Monte Carlo simulations. We also study how the 
non-equilibrium conditions influence the critical behaviour of 
the system. This problem is especially exciting because 
similar non-equilibrium perturbation results in a critical 
behaviour different from its equilibrium counterpart in some models where 
the global current vanishes \cite{schzia,sso}.
The finite-size scaling of the MC data indicates 
that the present model falls into the universality class of the 
equilibrium Ising model in agreement with a theoretical prediction \cite{grin}.

The next section is devoted to the violation of fluctuation-dissipation 
theorem (FDT) which is generally one of the benchmarks of non-equilibrium 
phenomena. The breakdown of this theorem results in a difference between 
the specific heats defined as the temperature 
derivative of the average energy on the one hand and derived
from the fluctuations of energy on the other hand. For example, in the uniformly 
driven model with attractive interaction the field suppresses the fluctuations 
and the difference is relevant \cite{zia,vm,ang}. Using both ways we calculate 
the specific heat for our model and show that the rate of the violation of FDT can 
be influenced by the variation of the fixed temperature. 

A wide variety of non-equilibrium systems exhibit "generic scale 
invariance", namely, long-range spatial correlations for arbitrary
parameter values. An example of such a system is the two-temperature
model with attractive interactions where a power law decay of pair 
correlation function is observed for all temperatures above criticality
\cite{cheng}. We also examine the 
two-point correlations of our model and find short-range correlations
at high temperatures.

\section{The model}

Consider a two-dimensional lattice gas on a square lattice with $L \times 
L = N$ sites under 
periodic boundary conditions. The occupation variables $n_i = 0$ or $1$ (if 
the lattice site is 
empty or occupied) and we assume half-filled occupation ($\sum_i n_i = N/2$). 
The
Hamiltonian is given by 
\begin{equation}
H = -J \sum_{(i,j)} n_i n_j  
\label{eq:hamilton}
\end{equation}
where the summation is over nearest-neighbour pairs and $J = -1$. Kawasaki 
exchange dynamics is 
assumed \cite{kaw}, namely the particles can jump to one of the empty 
nearest-neighbour sites. 
Anisotropic hopping rates are used for directions $\alpha = x$ or $y$:
\begin{equation}
g_\alpha (\Delta H) = { 1 \over {1 + \exp(\Delta H / T_\alpha)}}
\label{eq:hoppprob}
\end{equation}
where $\Delta H$ is the energy difference between the final and initial 
configurations. The 
lattice constant and the Boltzmann constant are chosen to be unity as usual.

In the case of $T_x = T_y$, the above model is equivalent to the celebrated 
kinetic Ising model which undergoes a continuous phase 
transition at $T_N = 0.567$. Below $T_N$ the system orders into a 
checkerboard-like pattern. 
If the lattice is divided into two interpenetrating 
sublattices ($A$ and $B$) then particles prefer staying in one of these sublattices. 
Hence the difference between the two averaged
sublattice occupations is the global order parameter: $m = 
| \rho_A - \rho_B |$, where $\rho_A$ denotes the averaged 
occupation of sublattice $A$. A suitable local order parameter
is the staggered magnetization at site $j$: $\phi_j = (-1)^j 
(2n_j - 1)$. It is important to note that $\phi$ is not 
conserved resulting in an important effect on the spatial correlations at high temperatures. 

When $T_x$ and $T_y$ are different the energy flows permanently from the 
heat bath of higher temperature to the lower one through the system and 
maintains a stationary non-equilibrium 
state. In other two-temperature model the previous studies set one of the 
temperatures (e.g. $T_y$) to be infinite, which reduced the number of 
control parameter to one \cite{cheng,pres} and the phase transition was 
studied as a function of $T_x$. Now we cannot use this simplification because the 
long-range ordered phase is expected to be unstable when one of the 
temperature is too high (the explanation of this expectation
will be given in the next section). Therefore we keep $T_y$ at a 
constant (finite) value when we vary $T_x$ to find the order-disorder transition point. 
Naturally similar behaviour would be obtained by exchanging the roles of 
the two temperatures.

\section{Phase transition and critical behaviour}

The study of non-equilibrium phase diagram by dynamical mean-field approach 
proved to be useful 
in many instances \cite{sso,dic}. In this approach we determine the 
stationary solution of the equations describing the time evolution of
configuration probabilities. At the level of $k$-point approximation, these
probabilities are given by probabilities of the possible configurations on $k$-point 
clusters. We refer to \cite{dic} for further details of this method.

\begin{figure}[htb]
\centerline{
       \epsfxsize=9.0cm
       \epsfbox{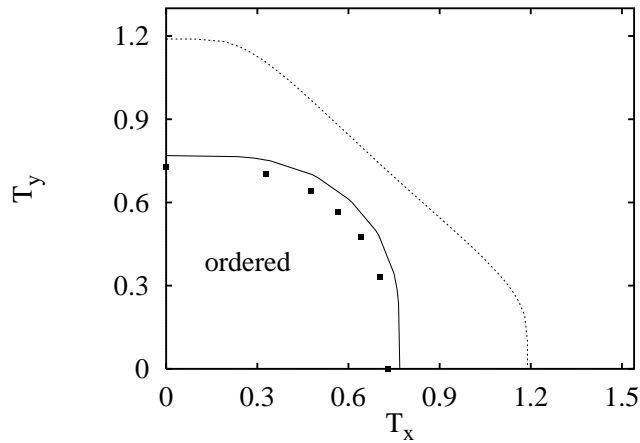}
               \vspace{0.5cm}
           }

\caption{Phase diagram predicted by two-point (dashed line) and
four-point (solid line) dynamical mean-field approximations. 
Filled boxes indicate the results of Monte Carlo simulation.}
\label{fig:phd}
\end{figure}

First we use the two-point level, 
consequently correlations larger than two lattice constants are neglected. 
In spite of this simplification, the simple
approximation predicts the relevant features of the phase diagram well as 
shown in Fig.~\ref{fig:phd}. Namely, the ordered
phase becomes unstable even at $T_y=0$ if $T_x$ is 
high enough. This behaviour may be understood qualitatively because
the nearest-neighbours of the occupied lattice sites are empty
in the ordered phase. Therefore the high temperature can 
influence the jumps and particles may leave the preferred
sublattice. The phase transition remains continuous at any 
value of $T_y$. Evidently, the quantitative predictions of this
approximation are far from the correct values
(e.g. $T_c^{2P}(T_x=T_y)=0.72, T_c^{2P}(T_y=0)=1.18$). 

\begin{figure}[htb]
\centerline{
       \epsfxsize=9.0cm
       \epsfbox{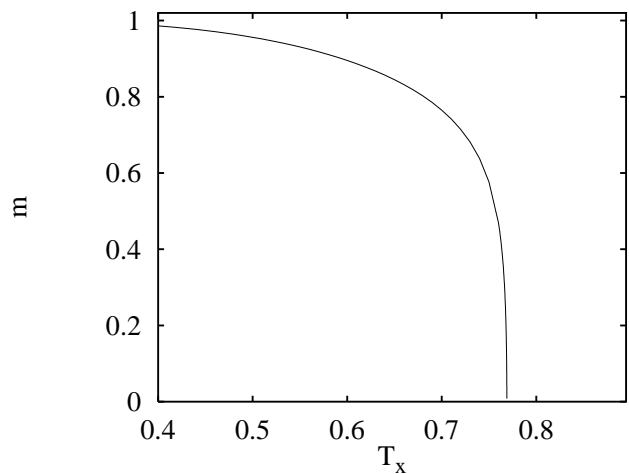}
               \vspace{0.5cm}
           }

\caption{Order parameter as a function of $T_x$ suggested by four-point 
approximation at $T_y = 0$ .}
\label{fig:op}
\end{figure}

To obtain more accurate data, we apply four-point approximation. Here we 
have to solve a system of eight coupled nonlinear equations numerically. The 
results (e.g. $T_c^{4P}(T_x=T_y)=0.60, T_c^{4P}(T_y=0)=0.76$) are
already comparable to the data of Monte Carlo simulation as shown in 
Fig.~\ref{fig:phd}. (The simulation data $T_c^{MC}(T_x=T_y)=0.56$ and 
$T_c^{MC}(T_y=0)=0.74)$). 
The phase transition is continuous even at $T_y = 0$, therefore the order 
parameter decreases gradually when $T_x$ tends to the
the transition point as Fig.~\ref{fig:op} shows. Naturally, we have the
same function when $T_x = 0$ and $T_y$ is varied. As a consequence the system 
becomes not perfectly ordered when we cool it ``horizontally'' to 
$T_x = 0$ axis on 
the phase diagram at a fixed nonzero value of $T_y$. 
Namely, the order parameter saturates to a value less than 1 
when $T_x \rightarrow 0$. 
The visualisation of particle distributions during the MC 
simulation shows that there are still some defects in the ordered 
homogenous phase even at $T_x = 0$ because some particles do
not stay in the preferred sublattice in consequence of the 
nonzero value of $T_y$.

The MC simulations were performed on a square lattice 
($L_x=L_y=L$) where system sizes ranged from $L = 20$ to 
$150$. At given ($T_x, T_y$) temperatures following a 
thermalization of $2,000 - 40,000$ MC steps, the time interval between 
two independent measurements ranged from $5$ 
MCS for $L=20$ systems to $100$ MCS for $L = 100$ systems. Increasing 
$T_x$ the whole procedure was repeated. To 
preclude the possibility of first-order transition, 
especially at low $T_y$, we repeated the measurement starting 
from high $T_x$ and decreasing $T_x$, but we have not 
detected hysteresis in the order parameter function. 
It is worth mentioning that the anisotropic dynamics does not 
result in anisotropy, namely the pair correlation functions 
to the $x$ and $y$ directions are identical within the limit 
of accuracy. 

We turn now to study the critical behaviour of the model.
For the previously studied two-temperature models the 
non-equilibrium behaviours differ from the equilibrium counterparts. In 
case of attractive interaction instead of the equilibrium Ising universality class, 
the anisotropic finite-size 
scaling resulted in new exponents $\beta = 0.33$ and $\nu = 0.60$ \cite{pres}. 
The other model with repulsive nearest- and next-nearest 
neighbour interactions leaves the equilibrium universality class 
of the x-y model 
with cubic anisotropy and is described by the Ising exponents \cite{sso}. 
Common features of these models are that the 
degenerate ground states violate the x-y symmetry of the systems.
For example in case of the attractive interaction
the particles condense into a strip oriented
either horizontally or vertically in the ordered phase.
In the other previously studied model with repulsive 
interactions the particles form alternately occupied and empty 
columns (or rows). The fact that the anisotropic dynamics breaks 
the x-y symmetry and reduces the number of the stable 
groundstates may result in a new non-equilibrium behaviour. In 
contrast to these models, in the present model the checkerboard-like groundstate 
does not break the x-y symmetry, therefore the non-equilibrium perturbation 
does not reduce the number of the possible groundstates.

As a consequence of the previous argument our conjecture is that the critical behaviour 
of the non-equilibrium model may be described by Ising exponents. This expectation 
agrees with the prediction of the field theoretic renormalisation group analysis 
\cite{zia,grin}.

\begin{figure}[htb]
\centerline{
       \epsfxsize=9.0cm
       \epsfbox{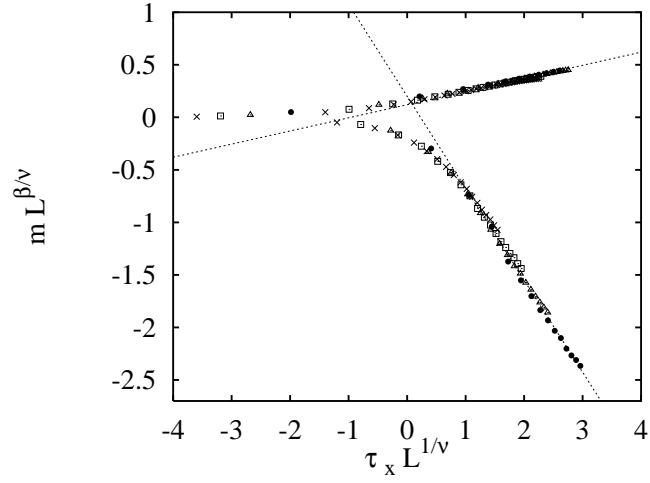}
               \vspace{0.5cm}
           }

\caption{Finite size scaling plots of MC data for the
order parameter at $T_y=0$. System sizes are $L = 20$ ($\times$),
$30$ ($\Box$), $50$ ($\bigtriangleup$), and $100$ ($\bullet$). 
The slopes of the inserted lines are $1/8$ and $-7/8$.}
\label{fig:fss}
\end{figure}

Although the dynamical mean-field approach provides a correct 
phase diagram, it is not applicable to 
describe critical behaviour. To study the critical behaviour
we have performed a finite size scaling of MC
data where the usual scaling form of the order 
parameter was used:   
\begin{equation} 
m(L,T_x) = L^{- \beta / \nu}\,\, \overline{m} ( \tau_x L^{1 / \nu} ) \,\,.
\label{eq:scale}
\end{equation}
Here $\tau_x$ denotes the reduced horizontal temperature 
deviation from the critical temperature.
Notice that $T_y$ is constant. Using $\beta = 1/8$ and
$\nu = 1$ in the scaling form, we have got nice data collapse
at $T_y = 0$ as Fig.~\ref{fig:fss} shows. Similar excellent data collapses
were found when we repeated the finite-size scaling at 
$T_y = 0.33$ and $0.64$. The Ising universality exponents
are believed to hold along the entire line of phase transitions.

\section{non-equilibrium fluctuations}

The character of the phase diagram makes possible to 
study the phase transition in a wide range of $T_y$. The
non-equilibrium phase transitions are generally 
accompanied by the violation of the FDT. It results in
difference between the specific heats defined as the 
derivative of energy with respect to the temperature ($C_D = \partial E /
\partial T$) and derived from the fluctuations in the energy
($C_F=(\langle E^2 \rangle - \langle E \rangle 
^2)/kT^2$, where $\langle \,\,\rangle$ stands for averaging
over time). To check this $C_D$ and $C_F$ are compared at 
different fixed values 
of $T_y$ when we cross the phase transition line by 
increasing $T_x$. In this case the correct definition of the 
two quantities are:

\begin{equation}
C_D = {\partial E \over \partial T_x} \bigg| _{T_y} 
\,\,\,\,\,\, \mbox{and} \,\,\,\, C_F = {\langle E^2 \rangle - \langle E \rangle 
^2 \over kT_{x}^2} \,\,.
\end{equation}

\begin{figure}[htb]
\centerline{
       \epsfxsize=9.0cm
       \epsfbox{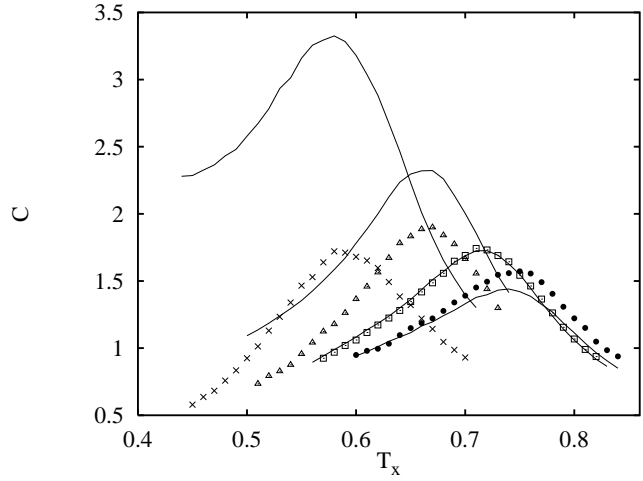}
               \vspace{0.5cm}
           }

\caption{MC data for specific heats at different values of $T_y$ ($0.64$
($\times$), $0.46$ ($\bigtriangleup$), $0.33$ ($\Box$), and $0$ ($\bullet$). Solid lines 
represent the specific heat calculated from the energy fluctuation.
The system sizes are $20 \times 20$. }
\label{fig:vfdt}
\end{figure}

At $T_y = 0$ the fluctuations are suppressed by the low temperature, 
therefore $C_F$ is smaller than $C_D$ even at the 
phase transition point. Increasing $T_y$, larger fluctuations 
can appear, which decrease the difference between $C_D$ and 
$C_F$. At a special value of $T_y = 0.33$,
the specific heats, which are calculated different ways, collapse. Further 
increasing $T_y$, the fluctuations become 
larger and $C_F$ exceeds $C_D$. Figure~\ref{fig:vfdt} 
indicates, how strongly the difference depends on $T_y$. Thus 
we can control the rate of violation of FDT by changing $T_y$, 
moreover, FDT seems to be valid at a special value of $T_y$. 
It is easy to detect that the $C_F$ curve is not symmetrical 
around the transition point ($T_c$) when $T_y = 0.64$. This feature
may be understood qualitatively. When $T_x$ is increased at fixed
value of $T_y$ near $0.64$, the system is close to the phase transition line
in the sub-critical region ($T_x < T_c$). While in the super-critical
region ($T_x < T_c$) the system is further away from the critical line.
Therefore the fluctuations are enhanced in the $T_x < T_c$ region comparing 
to the $T_x > T_c$ region.

\begin{figure}[htb]
\centerline{
       \epsfxsize=9.0cm
       \epsfbox{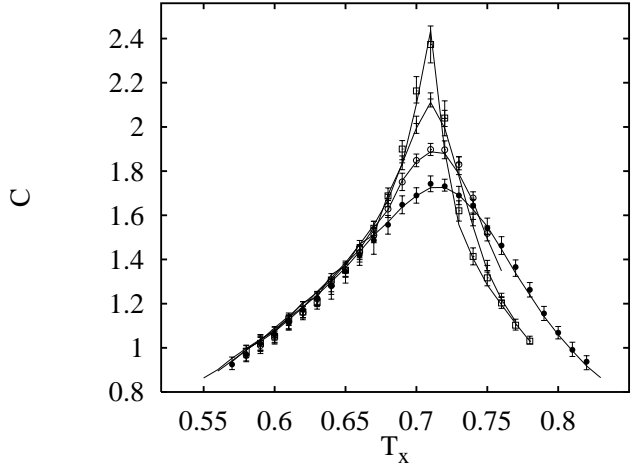}
               \vspace{0.5cm}
           }

\caption{Specific heats calculated both ways at $T_y = 0.33$ for 
different system sizes $L = 20$ ($\bullet$), $30$ ($\circ$), 
$50$ ($+$), and $100$ ( $\Box$). The error bars of $C_D$ are indicated.}
\label{fig:fdt}
\end{figure}

Though the data are obtained from small systems, similar 
behaviour can be observed for larger systems. This fact is 
illustrated in Fig.~\ref{fig:fdt} where the specific heats 
are plotted for different system sizes at the above mentioned 
special value of $T_y$. It is worth mentioning that in the
corresponding uniformly driven diffusive model the 
sharpening of the specific heat is not a monotonic function
of the system size \cite{ssab}.
The reason of this unusual behaviour is that the 
self-organising domain structure, which characterises the
thermodynamic limit, can only appear when the system sizes 
is large enough.  We expect traditional size dependence 
of specific heat in the two-temperature model because  
the reflection symmetric dynamics does not cause enhanced interfacial 
transport here. Although the peaks of the specific heats are more moderate
than in the equilibrium case, the intensity of the peaks increases with 
lattice size suggesting a divergence in the 
thermodynamic limit as Fig.~\ref{fig:fdt} shows.

Finally, we consider the spatial correlations above the 
phase transition point. Long-range correlations are so often observed 
in non-equilibrium systems that the exponential decay of spatial correlations 
seems to be exception. It is believed that in the non-equilibrium 
steady-state the spatially anisotropic conserved dynamics
produces such long-range correlations \cite{zhang,garr}. 
In agreement with this argument both simulations 
\cite{cheng} and perturbative approaches \cite{maes} found 
power law decay of correlations 
at high temperatures in different two-temperature models.
In our model the order parameter, which is the 'staggered' density, obeys 
no conservation law therefore the theory
predicts exponentially decaying correlations. However, the
question comes up whether the validity or the violation of
FDT may influence the spatial correlations somehow at high 
temperatures in the present model.

\begin{figure}[htb]
\centerline{
       \epsfxsize=9.0cm
       \epsfbox{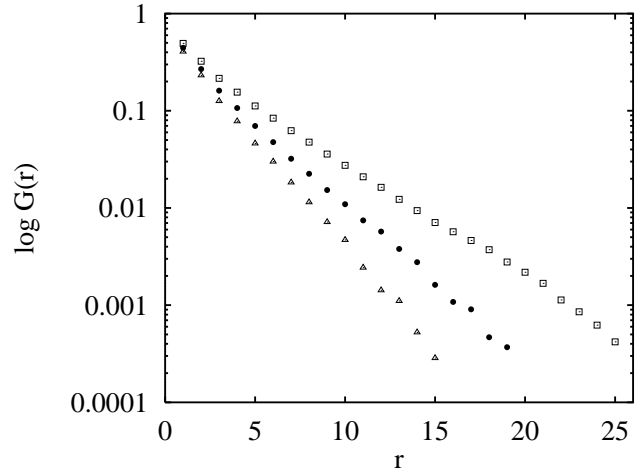}
               \vspace{0.5cm}
           }

\caption{Pair correlation functions at different $T_x$ above the critical 
region when $T_y = 0.33$ for all cases. $T_x = 1.2T_{c}$ ($\bigtriangleup$), $1.3T_{c}$ ($\bullet$), and $1.4T_{c}$ ($\Box$).}
\label{fig:corr}
\end{figure}

To measure the equal time pair-correlation function 
\begin{equation}
G(r)= \langle \phi(0) \phi (r) \rangle \,\,\,,
\end{equation}
where $\phi$ is the local order parameter, we chose the system 
size to be considerably larger than the 
correlation length. Thus we have chosen $L$ ranged from 50 to 
100 when we measured $G(r)$ at $T_y = 0.33$ for three 
different $T_x$ which are above the transition point ($T_x = 
1.2T_c, 1.3T_c,$ and $1.4T_c$). In every case $G(r)$ converged 
to zero. 
We have plotted $\log G(r)$ vs. $r$ in 
Fig.~\ref{fig:corr}. to demonstrate the 
exponential decay of correlations. Obviously the 
negative slopes become smaller when $T_x$ decreases
resulting in larger correlation lengths. Similar behaviours
were found for other values of $T_y$. 
  
\section{Conclusion}

The phase transition and critical behaviour of a 
two-temperature model with repulsive nearest-neighbour
interactions were studied by dynamical mean-field 
approximation and Monte Carlo simulations.
According to both approaches, the system undergoes a phase 
transition if one of the temperatures is high enough. The
phase transition is continuous even at zero fixed 
temperature and the critical behaviour is not affected by
the non-equilibrium perturbation. This result agrees with
the earlier theoretical expectations.

The present model gives an example that FDT can be valid or 
violate in case of the same non-equilibrium model at different 
values of the control parameter. The Monte Carlo data of the pair 
correlation function resulted exponential decay in the high 
temperature regime, which is in agreement with the 
theoretical prediction since the order parameter is not 
conserved. We found that the nature of spatial correlations 
are independent of the violation FDT.

\section{ ACKNOWLEDGEMENTS}
The author thank G. Szab\'o for his critical reading of the manuscript. He is grateful to Profs. Ole G. Mouritsen and 
Martin J. Zuckermann for helpful comments. It is pleasure to thank the hospitality of the Technical University of Denmark where this work was started. This research was supported by 
the Hungarian National Research Fund (OTKA) under Grant No. F-19560.

\vfill
\eject

\end{document}